\documentclass[aps,prl,twocolumn,superscriptaddress]{revtex4-1}
\usepackage{graphicx,latexsym}
\usepackage{amsmath,amssymb,amsfonts}
\usepackage{bm}
\usepackage{braket}
\usepackage{color}

\begin{document}

\title{Collective Excitations and Nonequilibrium Phase Transition in Dissipative Fermionic Superfluids}
\author{Kazuki Yamamoto}
\email{yamamoto.kazuki.72n@st.kyoto-u.ac.jp}
\affiliation{Department of Physics, Kyoto University, Kyoto 606-8502, Japan}
\author{Masaya Nakagawa}
\affiliation{Department of Physics, University of Tokyo, 7-3-1 Hongo, Tokyo 113-0033, Japan}
\author{Naoto Tsuji}
\affiliation{RIKEN Center for Emergent Matter Science (CEMS), Wako, Saitama 351-0198, Japan}
\affiliation{Department of Physics, University of Tokyo, 7-3-1 Hongo, Tokyo 113-0033, Japan}
\author{Masahito Ueda}
\affiliation{Department of Physics, University of Tokyo, 7-3-1 Hongo, Tokyo 113-0033, Japan}
\affiliation{RIKEN Center for Emergent Matter Science (CEMS), Wako, Saitama 351-0198, Japan}
\affiliation{Institute for Physics of Intelligence, University of Tokyo, 7-3-1 Hongo, Tokyo 113-0033, Japan}
\author{Norio Kawakami}
\affiliation{Department of Physics, Kyoto University, Kyoto 606-8502, Japan}

\date{\today}

\begin{abstract}
We predict a new mechanism to induce collective excitations and a nonequilibrium phase transition of fermionic superfluids via a sudden switch-on of two-body loss, for which we extend the BCS theory to fully incorporate a change in particle number. We find that a sudden switch-on of dissipation induces an amplitude oscillation of the superfluid order parameter accompanied by a chirped phase rotation as a consequence of particle loss. We demonstrate that when dissipation is introduced to one of the two superfluids coupled via a Josephson junction, it gives rise to a nonequilibrium dynamical phase transition characterized by the vanishing dc Josephson current. The dissipation-induced collective modes and nonequilibrium phase transition can be realized with ultracold fermionic atoms subject to inelastic collisions.
\end{abstract}

\maketitle

\textit{Introduction}.---Collective excitations of superconductors and superfluids have been widely studied in condensed matter physics \cite{Anderson58, Leggett66, Volkov74, Varma81, Varma82, Varma02, Andreev04, Barankov04, Barankov06A, Yuzbashyan062, Barankov06, Yuzbashyan06, Gurarie09, Yuzbashyan15, Hannibal15, Kett17, Hannibal18a, Hannibal18b, Pekker15, Shimano19}. Recent experimental progress in ultracold atoms has enabled studies of out-of-equilibrium dynamics of superfluids \cite{Hofs11, Bloch12, Kohl18, Kohl20}. For example, a periodic modulation of the amplitude of the order parameter excites the Higgs amplitude mode, which has been observed with ultracold fermions \cite{Kohl18} and in solid-state systems by light illumination on BCS superconductors \cite{Matsunaga13, Shimano14, Papen07, Papen08, Schnyder11, Zach13, Krull14, Tsuji15, Krull16}. As for collective phase modes, the Nambu-Goldstone mode exists in neutral superfluids, and the relative-phase Leggett mode has been predicted for multiband superfluids \cite{Leggett66, Sharapov02, Burnell10, Bittner15, Krull16, Cea16, Tsuji17}. In particular, ultracold atoms allow for a dynamical control of various system parameters, offering an ideal playground to investigate collective modes. However, they suffer from atom loss due to inelastic scattering, which has received little attention in literature.

In dissipative open quantum systems, the dynamics, after environmental degrees of freedom are traced out, is nonunitary and described by a completely positive and trace-preserving map \cite{Lindblad76, Daley14}. Such nonunitary dynamics is relevant for atomic, molecular, and optical systems, drastically changing various aspects of physics such as quantum critical phenomena \cite{Ashida17, Nakagawa18}, quantum phase transitions \cite{Diehl10, Honing12, Sieberer13}, quantum transport \cite{Damanet19, Yamamoto20} and superfluidity \cite{Han09, Yamamoto19}. In particular, high controllability of parameters in ultracold atoms has enabled investigations of non-equilibrium quantum dynamics induced by dissipation \cite{Tomita17, Ott13, Ott15, Ott16, Schneider17, Nagerl12, Spon18, Gerbier19, Tomita19, Takasu20, Witthaut08, Vidanovi14, Yamamoto19, Nakagawa20, Nakagawa21}, and studies of fermionic superfluidity in ultracold atoms undergoing inelastic collisions has achieved remarkable progress \cite{Han09, Zhang15, Iskin16, Xu16, He16, Hofer15, Pagano15, Cappellini19, Folling19, Yamamoto19}. The effect of particle loss in fermionic superfluids has been studied in the framework of the non-Hermitian BCS theory \cite{Yamamoto19}; however, it ignores a significant change in particle number due to quantum jumps. It is crucially important to go beyond the non-Hermitian framework to describe the long-time dynamics of a superfluid and associated collective modes of order parameters.

\begin{figure}[t]
\includegraphics[width=8.5cm]{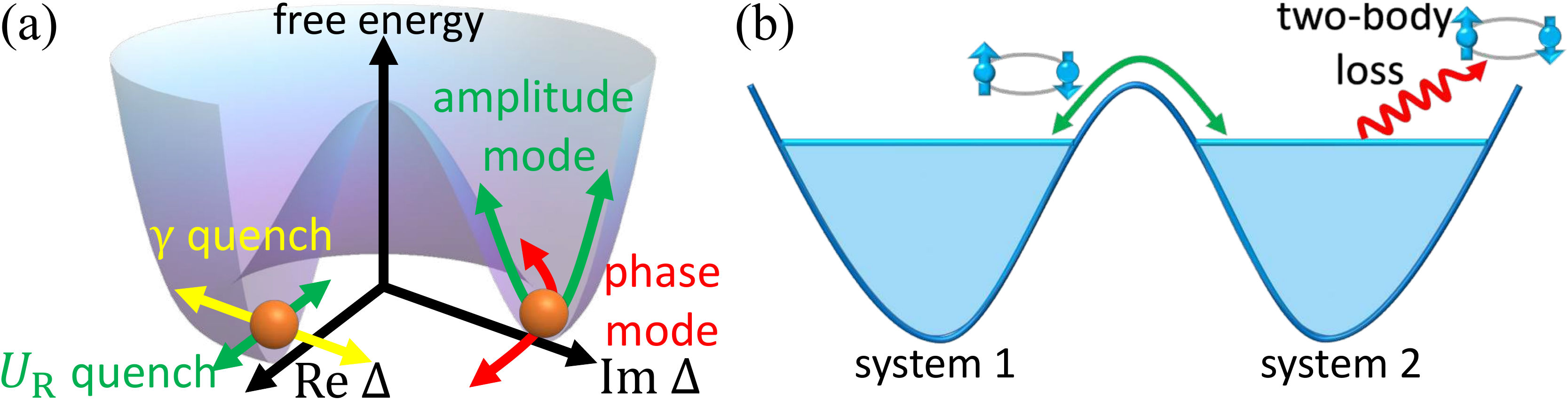}
\caption{(a) Schematic illustration of the amplitude and phase modes in a Mexican-hat free-energy potential as a function of the complex order parameter $\Delta$, when either the interaction $U_\mathrm{R}$ or the dissipation $\gamma$ is suddenly switched on. A sudden quench of the interaction $U_{\mathrm R}$ and that of the dissipation $\gamma$ kick $\Delta$ in a direction parallel and perpendicular to the radial direction, respectively. Note that a finite change of $\gamma$ excites both the phase and amplitude modes. (b) Two superfluids coupled via a Josephson junction, where one superfluid (system 2) is subject to two-body loss.}
\label{fig_Schematic}
\end{figure}

In this Letter, we theoretically investigate collective excitations and a nonequilibrium phase transition of fermionic superfluids driven by a sudden switch-on of two-particle loss due to inelastic collisions between atoms. By formulating a dissipative BCS theory that fully incorporates a change in particle number, we find that dissipation fundamentally alters the superfluid order parameter and induces collective oscillations in its amplitude and phase. In particular, we elucidate that a coupling between the order parameter and dissipation leads to a chirped phase rotation, in sharp contrast to the case of an interaction quench in closed systems [see Fig.~\ref{fig_Schematic}(a)].

To experimentally observe the collective phenomena induced by dissipation, we propose introducing a particle loss in one of two coupled superfluids to induce a relative-phase oscillation analogous to the Leggett mode \cite{Leggett66, Sharapov02, Burnell10, Bittner15, Krull16, Cea16, Tsuji17} [see Fig.~\ref{fig_Schematic}(b)]. The phase mode causes an oscillation of a Josephson current around a nonvanishing dc component. Remarkably, when dissipation becomes strong, the coupled system undergoes a nonequilibrium phase transition characterized by the vanishing dc Josephson current, which can be regarded as a generalization of a dynamical phase transition \cite{Barankov06, Yuzbashyan06, Thywissen19, Muniz20} to dissipative quantum systems. Our findings can experimentally be tested with ultracold atoms through introduction of dissipation via a photoassociation process \cite{Tomita17, Takasu20}.


\textit{Dissipative BCS theory}.---We consider ultracold fermionic atoms described by the three-dimensional attractive Hubbard model
\begin{align}
H=\sum_{\bm{k}\sigma}\epsilon_{\bm{k}}c_{\bm{k}\sigma}^\dagger{c}_{\bm{k}\sigma}-U_{\mathrm{R}}\sum_{i}c_{i\uparrow}^{\dagger}c_{i\downarrow}^{\dagger}c_{i\downarrow}c_{i\uparrow},
\label{eq_Hubbard}
\end{align}
where $U_{\mathrm{R}}>0$, $\epsilon_{\bm{k}}$ is the single-particle energy dispersion, and $c_{\bm k\sigma}$ ($c_{i\sigma}$) denotes the annihilation operator of a spin-$\sigma$ fermion with momentum $\bm{k}$ (at site $i$). When the system is subject to inelastic collisions, scattered atoms are lost to a surrounding environment, resulting in dissipative dynamics as observed experimentally \cite{Tomita17, Nagerl12, Spon18, Tomita19}. Here, we study the time evolution of the density matrix $\rho$ which is described by the Lindblad equation \cite{Lindblad76, Daley14}
\begin{align}
\frac{d\rho}{dt}=\mathcal{L}\rho=-i[H,\rho]-\frac{\gamma}{2}\sum_i (\{L_i^\dag {L}_i, \rho\} -2L_i \rho L_i^\dagger),
\label{eq_Lindblad}
\end{align}
where $L_i = c_{i\downarrow} c_{i \uparrow}$ is a Lindblad operator that describes two-body loss with loss rate $\gamma>0$. We note that the kinetic energy of lost atoms is large because of large internal energy of atoms before inelastic collisions. Under such situations, atoms after inelastic collisions are quickly lost into the surrounding environment and the Born-Markov approximation is justified \cite{Syassen08, Garcia09, Durr09}.

We first study how the standard BCS theory is generalized in open dissipative systems by formulating a time-dependent mean-field theory in terms of a closed-time-contour path integral \cite{Diehl16, Kamenev11}. We start with a generating functional defined as
\begin{align}
Z=\mathrm{tr}\rho=\int\mathcal D[c_-, \bar c_-, c_+, \bar c_+]e^{iS}=1,\label{eq_pf}
\end{align}
with an action
\begin{align}
&S=\int_{-\infty}^\infty dt \Big[\sum_{\bm k \sigma}(\bar c_{\bm k \sigma +} i\partial_t c_{\bm k \sigma +} -\bar c_{\bm k \sigma -}i\partial_t c_{\bm k \sigma -})-H_+\notag\\
&+ H_- + \frac{i \gamma}{2}\sum_i(\bar L_{i+}L_{i+}+\bar L_{i-}L_{i-}-2L_{i+}\bar L_{i-})\Big],
\label{eq_action}
\end{align}
where the subscripts $+$ and $-$ denote forward and backward paths, $H_\alpha= \sum_{\bm{k}\sigma}\epsilon_{\bm{k}}\bar c_{\bm{k}\sigma\alpha} c_{\bm{k}\sigma\alpha}-U_{\mathrm{R}}\sum_{i}\bar c_{i\uparrow\alpha}\bar c_{i\downarrow\alpha} c_{i\downarrow\alpha}c_{i\uparrow\alpha}$, $L_{i\alpha}=c_{i\downarrow\alpha} c_{i \uparrow\alpha}$, and $\bar L_{i\alpha}=\bar c_{i \uparrow\alpha}\bar c_{i\downarrow\alpha}$ $(\alpha=+,-)$. Note that the action has U(1) symmetry under $c_{i\sigma\alpha}\to e^{i\theta}c_{i\sigma\alpha}$ though the particle number is not conserved \cite{Buca12, Albert14}. By introducing auxiliary fields via the Hubbard-Stratonovich transformation, we rewrite the action in a quadratic form of fermionic Grassmann fields as  \cite{Yamamoto19, Supple}
\begin{align}
S=\int dt\Big\{&\sum_{\bm k}\Big[
\bar \psi_{\bm k +}^t
\left(
\begin{matrix}
i\partial_t-\epsilon_{\bm{k}}&-\Delta\\
-\Delta^*&-i\partial_t+\epsilon_{\bm k}
\end{matrix}
\right)
\psi_{\bm k +}\notag\\
&-
\bar \psi_{\bm k -}^t
\left(
\begin{matrix}
i\partial_t-\epsilon_{\bm k}&-\Delta\\
-\Delta^*&-i\partial_t+\epsilon_{\bm k}
\end{matrix}
\right)
\psi_{\bm k -}\Big]\Big\},
\label{eq_action2}
\end{align}
where $\bar \psi_{\bm k \alpha}=(
\begin{matrix}
\bar c_{\bm{k}\uparrow\alpha},&c_{-\bm{k}\downarrow\alpha}
\end{matrix}
)^t$ and $\psi_{\bm k \alpha}=(
\begin{matrix}
c_{\bm{k}\uparrow\alpha},&\bar c_{-\bm{k}\downarrow\alpha}
\end{matrix}
)^t$ $(\alpha=+, -)$. Here $\Delta$ is the superfluid order parameter which can be determined from the requirement that the action be extremal as \cite{Supple}
\begin{align}
\Delta=-\frac{U}{N_0}\sum_{\bm k}\mathrm{tr}(c_{-\bm k\downarrow}c_{\bm k \uparrow}\rho)\equiv-\frac{U}{N_0}\sum_{\bm k}\langle c_{-\bm k\downarrow}c_{\bm k \uparrow}\rangle,
\end{align}
where $U=U_R+i\gamma/2$ is an effective complex coupling constant including a contribution from the atom loss \cite{Yamamoto19}, and $N_0$ is the number of sites. Importantly, the order parameter includes the loss rate $\gamma$, which leads to dissipation-induced collective modes as discussed below. The action \eqref{eq_action2} describes the mean-field time-evolution equation of the density matrix as
\begin{gather}
\frac{d\rho}{dt}=-i[H_{\mathrm{eff}},\rho],
\label{eq_unitary}\\
H_{\mathrm{eff}}=\sum_{\bm{k}}
\Psi_{\bm k}^\dag
\left(
\begin{matrix}
\epsilon_{\bm{k}}&\Delta\\
\Delta^*&-\epsilon_{\bm{k}}
\end{matrix}
\right)
\Psi_{\bm k},
\label{eq_HamiltonianEff}
\end{gather}
where $\Psi_{\bm k}=(\begin{matrix}c_{\bm k \uparrow},&c_{-\bm k \downarrow}^\dag \end{matrix})^t$ is the Nambu spinor. In the Supplemental Material \cite{Supple}, we show that Eq.~\eqref{eq_unitary} can be derived from two different methods, i.e. the mean-field theory for the Lindblad equation and the time-dependent Bogoliubov-de Gennes analysis. While Eq.~\eqref{eq_unitary} appears to describe unitary evolution, it is consistent with the original Lindblad equation \eqref{eq_Lindblad} as a consequence of the time-dependent BCS ansatz \cite{Supple}.

We use Anderson's pseudospin representation \cite{Anderson58, Barankov04, Tsuji15, Barankov06A, Yuzbashyan062, Yuzbashyan06, Yuzbashyan15, Andreev04, Barankov06} defined by $\bm \sigma_{\bm k}=\frac{1}{2}\Psi_{\bm k}^\dag\cdot\bm \tau\cdot\Psi_{\bm k}$ and $H_{\mathrm{eff}}=2\sum_{\bm k}\bm b_{\bm k}\cdot\bm \sigma_{\bm k}$, where $\bm \tau=(\begin{matrix}\tau_x,&\tau_y,&\tau_z\end{matrix})$ is the vector of the Pauli matrices. The pseudospins satisfy the commutation relations $[\sigma_{\bm k}^j,\sigma_{\bm k}^k]=i\epsilon_{jkl}\sigma_{\bm k}^l$. For simplicity of notation, we omit the bracket and regard $\bm \sigma_{\bm k}$ as the expectation value of the pseudospin operator. By using the commutation relation of the pseudospins, Eq.~\eqref{eq_unitary} is mapped to the Bloch equation:
\begin{gather}
\frac{d\bm \sigma_{\bm k}}{dt}
=2\bm b_{\bm k}\times\bm \sigma_{\bm k},\label{eq_bloch}\\
\bm b_{\bm k}=(\begin{matrix}\mathrm{Re}\Delta,&-\mathrm{Im}\Delta,&\epsilon_{\bm k}\end{matrix}).
\end{gather}
Equation~\eqref{eq_bloch} shows that the superfluid dynamics is characterized by precession of a pseudospin in an effective magnetic field $\bm b_k$. Here, the order parameter is determined self-consistently from the pseudospin expectation value as
\begin{align}
\Delta=|\Delta|e^{i\theta}=-\frac{U}{N_0}\sum_{\bm k}(\sigma_{\bm k}^x-i\sigma_{\bm k}^y).
\label{eq_order}
\end{align}
It is noteworthy that the norm of the pseudospin is conserved by the Bloch equation \eqref{eq_bloch}. The time evolution of the particle number due to particle loss is obtained from Eq.~\eqref{eq_unitary} as
\begin{align}
\frac{1}{N_0}\frac{dN}{dt}=-\frac{2\gamma|\Delta|^2}{|U|^2},
\label{eq_Ndt}
\end{align}
which reflects the dynamics of the order parameter.

\begin{figure}[t]
\includegraphics[width=8.5cm]{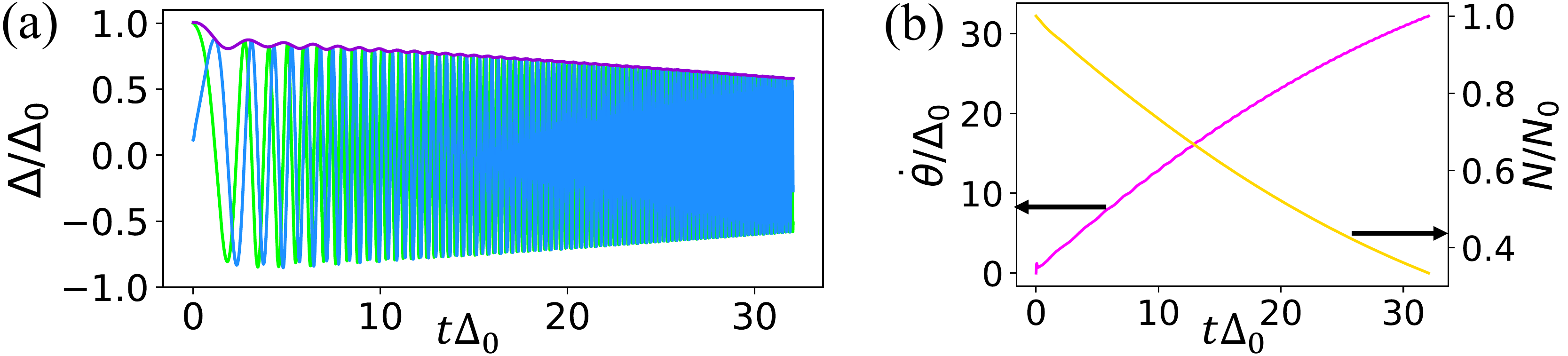}
\caption{Dynamics of a superfluid after the atom loss with $\gamma=2.81\Delta_0$ is switched on for the initial state with $U_{\mathrm{R}}=12.2\Delta_0$  and bandwidth $W=46.8\Delta_0$, where $\Delta_0$ is the superfluid order parameter in the absence of the atom loss. (a) Real parts (light green), imaginary parts (blue), and the amplitude (violet) of the order parameter. (b) Angular velocity (pink) and particle number (yellow) plotted against time. The figures indicate a chirped phase rotation and an amplitude oscillation of $\Delta$.}
\label{fig_lossquench}
\end{figure}

\textit{Collective excitations: phase and amplitude modes}.---We numerically solve the Bloch equation \eqref{eq_bloch} self-consistently under the condition \eqref{eq_order}. As an initial state, we prepare a BCS ground state with $\gamma=0$, whose pseudospin representation is given by $\sigma_{\bm k}^x(0)=-\Delta_0/\sqrt{{\epsilon_{\bm k}}^2+\Delta_0^2}$, $\sigma_{\bm k}^y(0)=0$ and $\sigma_{\bm k}^z(0)=-\epsilon_{\bm k}/\sqrt{{\epsilon_{\bm k}}^2+\Delta_0^2}$ with $\Delta_0\in\mathbb{R}$. The single-particle energy $\epsilon_{\bm k}$ is measured from the Fermi energy of the initial state. The bandwidth $W$ is defined by the energy difference between the upper and lower edges of the energy spectrum with a constant density of states. We then switch on the atom loss $\gamma$ at $t=0$. The results shown in Fig.~\ref{fig_lossquench} are obtained by the second-order Runge-Kutta method.
In the long-time limit, the amplitude of the superfluid order parameter $\Delta$ is suppressed due to dissipation, indicating a decay of superfluidity [see Fig.~\ref{fig_lossquench}(a)]. We note that the order parameter decays in the long-time limit due to a decrease of the particle number [see Fig.~\ref{fig_lossquench}(b)], and such behavior has no counterpart in the quench in isolated systems \cite{Barankov06, Yuzbashyan062}. Remarkably, after the dissipation $\gamma$ is introduced, the U(1) phase of the order parameter rotates and shows chirping, i.e., its angular velocity increases with time [see Fig.~\ref{fig_lossquench}(a), (b)] as a consequence of the dynamical shift of the Fermi level \cite{Supple}. This property is unique to the dissipative superfluid and distinct from the usual dynamics in isolated systems where the U(1) phase stays constant \cite{Yuzbashyan062, Barankov06, Yuzbashyan06}. The phase rotation is understood from an initial-state free energy as a function of $\Delta$ [see Fig.~\ref{fig_Schematic}(a)]. When dissipation is introduced, the sudden quench of the imaginary part of $U$ in Eq.~\eqref{eq_order} pushes the order parameter towards the direction perpendicular to the radial direction irrespective of the initial choice of the gauge. Another way to understand the phase rotation is to introduce an effective chemical potential as $\Delta(t)=\exp(-2i\int_0^t \mu_{\mathrm{eff}}(t)dt)\Omega(t)$ $(\Omega\in\mathbb R)$. By performing a global gauge transformation from $\Delta$ to $\Omega$, the Bloch equation is written in the Larmor frame on which the energy dispersion is given by $\xi_{\bm k}(t) = \epsilon_{\bm k} - \mu_{\mathrm{eff}}(t)$. This gauge transformation indicates that the phase rotation corresponds to a decrease of the effective chemical potential, which is consistent with the behaviors of $\dot \theta$ and $N$ in Fig.~\ref{fig_lossquench}(b). This result can naturally be understood since the phase and the particle number are conjugate variables.

We also find amplitude oscillations in $|\Delta|$ as shown in Fig.~\ref{fig_lossquench}(a). The amplitude oscillations are more pronounced when the interaction and the dissipation are simultaneously quenched \cite{Supple}. The mechanism behind the oscillations is that the quench of the imaginary part of $U$ changes the absolute value of $\Delta$ [see Fig.~\ref{fig_Schematic}(a)]. The frequency of the amplitude oscillation is close to $2\Delta_0$ at an early stage, and increases as time evolves. This behavior is distinct from that of an isolated system, where the amplitude mode is characterized by the constant frequency. Such behavior can be observed from the measurement of the time-dependent particle number via Eq.~\eqref{eq_Ndt}.
\begin{figure*}[t]
\includegraphics[width=18cm]{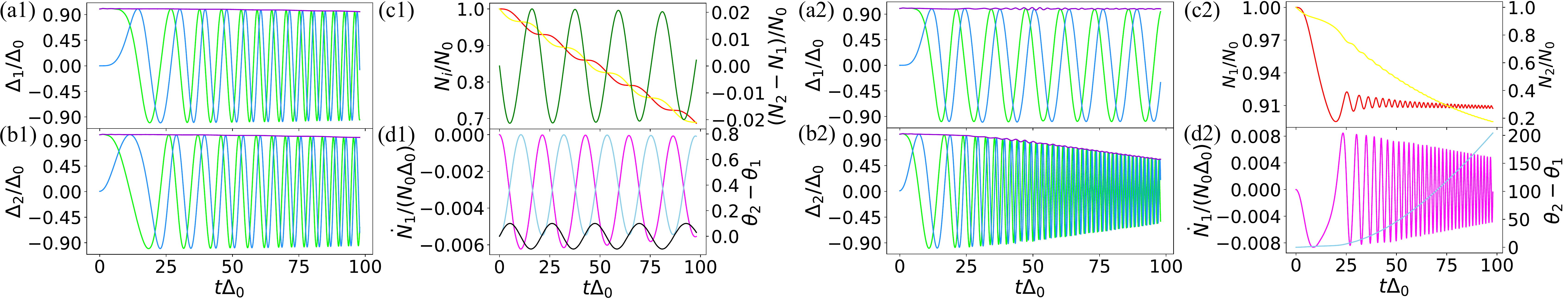}
\caption{Dynamics of two fermionic superfluids after the switch-on of the atom loss $\gamma$ and the tunnel coupling $V=0.02\Delta_0$ with $U_{\mathrm R}=3.06\Delta_0$ and bandwidth $W=5.11\Delta_0$, where $\gamma=0.03\Delta_0$ for (a1)-(d1) and $\gamma=0.06\Delta_0$ for (a2)-(d2). (a), (b) Real parts (light green), imaginary parts (blue), and amplitudes (violet) of the order parameter for systems 1 and 2. (c) Particle numbers of system 1 (red) and system 2 (yellow), and their difference [green, in (c1)]. (d) Josephson current (pink) and phase difference (light blue) between the two systems. The black curve in (d1) shows an oscillation at frequency $\omega_\mathrm{L}$ for comparison.}
\label{fig_multiquench}
\end{figure*}

\textit{Collective excitations: Leggett mode}.---To observe the chirped phase rotation of the superfluid order parameter that is a unique feature of dissipative superfluids, we propose that the phase rotation induced by dissipation can be detected when two superfluids are connected via a Josephson junction, which has been realized in ultracold atoms \cite{Spun07, Roati15, Roati18, Moritz20, Roati20}. As the phase difference in the two superfluid order parameters is gauge-invariant, it leads to an observable Josephson current. We introduce dissipation to one of the two superfluids as schematically illustrated in Fig.~\ref{fig_Schematic}(b) and assume that they are coupled via a tunneling Hamiltonian \cite{Tsuji17, Leggett66}
\begin{align}
H_{\mathrm{tun}}=-\frac{V}{N_0}\sum_{\bm k \bm{k}'}( c_{1\bm{k} \uparrow}^\dag c_{1-\bm{k} \downarrow}^\dag c_{2-\bm k'\downarrow}c_{2\bm k' \uparrow}+\mathrm{H.c.}),
\label{eq_tunnel}
\end{align}
where $V>0$ is the amplitude of Cooper-pair tunneling between system $1$ without dissipation and system $2$ with two-particle loss. By performing a mean-field analysis, we can write the system Hamiltonian as $H_{\mathrm{sys}} =H_1 + H_2 + H_{\mathrm{tun}} = H_1^\prime+ H_2^\prime$, where $H_i\equiv\sum_{\bm k \sigma}\epsilon_{\bm k} c_{i\bm k \sigma}^\dag c_{i\bm k \sigma} + \sum_{\bm k}(\Delta_i c_{i \bm k \uparrow}^\dag c_{i -\bm k \downarrow}^\dag + \mathrm{H.c.})$ ($i=1,2$) is the mean-field Hamiltonian of system $i$ and $H_i^\prime \equiv H_i  -V/N_0\sum_{\bm k \bm k'} (\langle c_{j -\bm k'\downarrow}c_{j\bm k' \uparrow}\rangle c_{i\bm k \uparrow}^\dag c_{i-\bm k \downarrow}^\dag + \mathrm{H.c.})$ [$(i, j) = (1, 2)$ or $(2, 1)$]. In the pseudospin respresentation, the Hamiltonian is written as
$H_i^\prime=2\sum_{\bm k}\bm b_{i\bm k}\cdot\bm \sigma_{i\bm k}$ with an effective magnetic field $\bm b_{i\bm k}
=(
\begin{matrix}
\mathrm{Re}\Delta_i^\prime,&-\mathrm{Im}\Delta_i^\prime,&\epsilon_{i\bm k}
\end{matrix}
)$, 
which yields the Bloch equation $d\bm \sigma_{i\bm k}/dt=2\bm b_{i\bm k}\times\bm \sigma_{i\bm k}$. The self-consistent conditions for the order parameters read $\Delta_1
=|\Delta_1|e^{i\theta_1}=-\frac{U_{\mathrm{R}}}{N_0}\sum_{\bm k}(\sigma_{1 \bm k}^x-i\sigma_{1 \bm k}^y)$ and
$\Delta_2
=|\Delta_2|e^{i\theta_2}=-\frac{U}{N_0}\sum_{\bm k}(\sigma_{2 \bm k}^x-i\sigma_{2 \bm k}^y)$,
where $N_0$ is the number of sites of each system. Here, the relations $\Delta_i^\prime=\Delta_i-V/N_0\sum_{\bm k}(\sigma_{j \bm k}^x-i\sigma_{j \bm k}^y)$ [$(i, j) = (1, 2)$ or $(2, 1)$] are satisfied. Then, the Josephson current between the two superfluids is given by the rate of change in the particle number of system 1:
\begin{align}
\frac{1}{N_0}\frac{dN_1}{dt}
=-\frac{4V|\Delta_1||\Delta_2|}{U_{\mathrm{R}}|U|}\sin\left(\theta_2-\theta_1+\delta\right),
\label{eq_JosephsonCurrent}
\end{align}
where $\delta=\tan^{-1}(-\gamma/2U_{\mathrm{R}})$ is the phase shift due to the sudden switch-on of the atom loss.

We numerically solve the coupled Bloch equations for $\bm \sigma_{i\bm k}$. We assume that dissipation $\gamma$ and tunneling $V$ are turned on at $t=0$ for the BCS ground state. The numerical results for weak dissipation are shown in Fig.~\ref{fig_multiquench}(a1)-(d1). In Figs.~\ref{fig_multiquench}(a1) and (b1), the dynamics of two superfluids almost synchronize with each other because the time scale of particle loss is comparable with the inverse tunneling rate. In the pseudospin picture, the dynamics of particle numbers shown in Fig.~\ref{fig_multiquench}(c1) can be interpreted as the nutation of pseudospins. Importantly, we see that, although the particle number of the system decreases in time, the corresponding amplitude of the order parameter stays almost constant. This implies that the condensate fraction against the total particle number becomes larger than that of the initial state. As inferred from Fig.~\ref{fig_multiquench}(d1), the Josephson current oscillates around its dc component. Such behavior is reminiscent of Shapiro steps in a Josephson junction under irradiation of a microwave \cite{Tinkham04}; however, in the present case, the Josephson current oscillate \textit{spontaneously without any external field}. Moreover, from Fig.~\ref{fig_multiquench}(d1), the frequency of the oscillation of the phase difference between the two systems is close to that of the relative-phase mode known as the Leggett mode \cite{Leggett66, Tsuji17} whose dispersion relation is given by $\omega_{\mathrm L}=2\sqrt{(\lambda_{12}+\lambda_{21})|\Delta_1||\Delta_2|/\mathrm{det}\lambda}$, where $\lambda_{11}=\lambda_{22}=U_{\mathrm{R}}/W$, $\lambda_{12}=\lambda_{21}=V/W$ and $\mathrm{det} \lambda=\lambda_{11}\lambda_{22}-\lambda_{12}\lambda_{21}$. We note that $\omega_L$ includes the effect of loss through the order parameters. The Leggett mode with frequency $\omega_{\mathrm L}$ has been discussed in the context of a collective mode in a multiband superconductor irradiated by light \cite{Tsuji17}. The agreement between the frequencies of the relative-phase modes in very different situations can be understood as follows. When dissipation is weak, the time evolution of an order parameter is given by $\Delta_i(t) = \exp(-2i\int_0^t dt\mu_{i\mathrm{eff}}(t))|\Delta_i(t)|$ with an effective chemical potentials $\mu_{i\mathrm{eff}}$. Then, by performing a global gauge transformation from $c_{i\bm k \sigma}$ to $c_{i\bm k \sigma} \exp(i\int_0^t\sum_i\mu_{i\mathrm {eff}}dt/2)$, we can linearize the Bloch equation with respect to the relative phase difference between $\Delta_i$'s by following Ref.~\cite{Tsuji17}.
\begin{figure}[b]
\includegraphics[width=8.5cm]{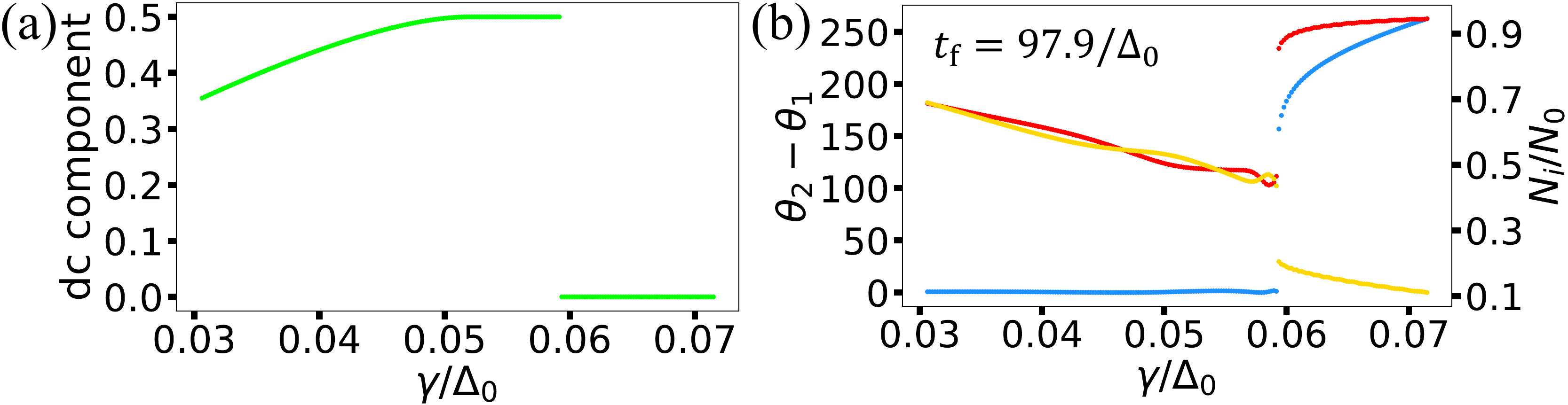}
\caption{(a) DC component of the Josephson oscillation defined by ($\max_{0\le t \le t_{\mathrm f}}\{\sin\left(\theta_2(t)-\theta_1(t)+\delta\right)\} + \min_{0\le t \le t_{\mathrm f}}\{\sin\left(\theta_2(t)-\theta_1(t)+\delta\right)\})/2$ with $t_{\mathrm f} = 97.9 / \Delta_0$. (b) Phase difference between the two systems (blue) and particle numbers of system 1 (red) and system 2 (yellow) after a sufficiently long time ($t_{\mathrm f}=97.9/\Delta_0$). The parameters used are $U_{\mathrm{R}}=3.06\Delta_0$, $V=0.02\Delta_0$, and $W=5.11\Delta_0$.}
\label{fig_transition}
\end{figure}

\textit{Nonequilibrium phase transition}.---In the presence of strong dissipation, the order parameter of system 2 oscillates faster than that of system 1 [see Fig.~\ref{fig_multiquench}(a2), (b2)] and the phase difference monotonically increases in time [see Fig.~\ref{fig_multiquench}(d2)]. This is because the dissipation rate larger than the tunneling rate makes system 1 fail to follow the decay of system 2, resulting in the dynamics similar to that of a single superfluid shown in Fig.~\ref{fig_lossquench}. In particular, the chirped phase rotation of the superfluid order parameter of system 2 can be detected from the Josephson current [Fig.~\ref{fig_multiquench}(d2)]. As the superfluidity of system 2 is suppressed, the Josephson current also decays, and the particle number of system 1 settles to a constant after some transient time [see Fig.~\ref{fig_multiquench}(c2)]. The latter behavior is attributed to the continuous quantum Zeno effect \cite{Syassen08, Garcia09, Zhu14, Daley09, Yan13, Yamamoto19}, which states that strong dissipation prevents tunneling and inhibits loss in system 1. In fact, an effective decay rate of system 1 is given by $\gamma_{\mathrm{eff}} \equiv |V_{\mathrm{eff}}|^2 / \gamma$ with an effective tunneling rate $V_{\mathrm{eff}} = V\Delta_2 / U_{\mathrm R}$ from Eq.~\eqref{eq_tunnel}, leading to suppression of decay $\gamma_{\mathrm{eff}} \to 0$ for $|\Delta_2|^2/\gamma \to 0$.

The two dynamically distinct regimes of superfluid behaviors suggest the existence of dynamical phases of matter \cite{Thywissen19, Muniz20} in dissipative superfluids. The qualitative change in the superfluid behaviors with respect to the dissipation strength highlights a dynamical phase transition characterized by the vanishing dc Josephson current [Fig.~\ref{fig_transition}(a)], where the dc component of the Josephson oscillation is defined by ($\max_{0\le t \le t_{\mathrm f}}\{\sin\left(\theta_2(t)-\theta_1(t)+\delta\right)\} + \min_{0\le t \le t_{\mathrm f}}\{\sin\left(\theta_2(t)-\theta_1(t)+\delta\right)\})/2$ [see Eq.~\eqref{eq_JosephsonCurrent}] after a sufficiently long time evolution with $t_{\mathrm f} = 97.9 / \Delta_0$. We emphasize that the dynamical phase transition in dissipative superfluids is essentially distinct from the phase transition between ground states in a non-Hermitian BCS superfluid \cite{Yamamoto19}. The former is caused by a change in particle number in the long-time dynamics, whereas the latter is caused by an exceptional point of a non-Hermitian BCS Hamiltonian, which is relevant to the short-time dynamics during which the number of particles does not change \cite{Footnote}. From Fig.~\ref{fig_transition}(b), we see that the phase difference $\theta_2-\theta_1$ starts to increase monotonically at the critical point and that the difference in particle number $(N_2-N_1)/N_0$ becomes much larger. The behavior of the phase difference is reminiscent of the localization-diffusion transition of a quantum-mechanical particle moving in a washboard potential in the presence of frictional force \cite{Caldeira81, Schdmit83, Guinea85}. However, the origin of the transition shown in Fig.~\ref{fig_transition} is essentially different from frictional force, since it cannot change the particle number. In fact, as shown in the Supplemental Material \cite{Supple}, the dynamical phase transition in Fig.~\ref{fig_transition} is triggered by the competition between the Josephson coupling and particle loss. Moreover, as the steady state is a vacuum due to the particle loss, the dynamical phase transition is observed only in the transient dynamics, and thus distinct from steady-state transitions.

\textit{Conclusions.}---We have investigated the loss-quench dynamics of fermionic superfluids, and have demonstrated that the dynamics exhibits amplitude and phase modes with chirped oscillations, the latter of which is a salient feature of a dissipative superfluid. To observe the chirped phase rotation, we have proposed a Josephson junction comprised of dissipative and nondissipative superfluids. We have shown that the relative-phase Leggett mode can be detected from the Josephson current for weak dissipation. Remarkably, when dissipation becomes strong, the superfluids exhibit the unique nonequilibrium phase transition triggered by particle loss. Our prediction can be tested with ultracold atomic systems of $^6$Li  \cite{Roati15, Roati18}, for example, by introducing dissipation using photoassociation processes \cite{Tomita17, Takasu20}. It is of interest to explore how the dimensionality or confinement by a trap potential affects the dynamics and associated collective modes \cite{Hannibal15, Kett17, Hannibal18a, Hannibal18b}.

\begin{acknowledgments}
We are grateful to Yuto Ashida, Philipp Werner, Shuntaro Sumita, and Yoshiro Takahashi for fruitful discussions. This work was supported by KAKENHI (Grants No.\ JP18H01140, No.\ JP18H01145, and No. JP19H01838) and a Grant-in-Aid for Scientific Research on Innovative Areas (KAKENHI Grant No.\ JP15H05855) from the Japan Society for the Promotion of Science. K.Y. was supported by WISE Program, MEXT and JSPS KAKENHI Grant-in-Aid for JSPS fellows Grant No. JP20J21318. M.N. was supported by KAKENHI (Grant No. JP20K14383). N.T. acknowledges support by JST PRESTO (Grant No. JPMJPR16N7) and KAKENHI (Grant No. JP20K03811).
\end{acknowledgments}

\nocite{apsrev41Control}
\bibliographystyle{apsrev4-1}
\bibliography{OpenBCS.bib}

\begin{thebibliography}{97}
\bibitem [{Foo()}]{Footnote}%
  \BibitemOpen
  \href@noop {} {}\bibinfo {note} {In Ref.~\cite{Nakagawa21}, it is shown that
  exceptional points of a non-Hermitian Hamiltonian and a Liouvillian are the
  same when the system has either gain or loss. Thus, the Liouvillian under
  consideration exhibits the spectral singularity originating from the same
  exceptional points as those of the non-Hermitian BCS Hamiltonian
  \cite{Yamamoto19}.}\BibitemShut {Stop}%
\end{thebibliography}


\clearpage

\renewcommand{\thesection}{S\arabic{section}}
\renewcommand{\theequation}{S\arabic{equation}}
\setcounter{equation}{0}
\renewcommand{\thefigure}{S\arabic{figure}}
\setcounter{figure}{0}

\onecolumngrid
\appendix
\begin{center}
\large{Supplemental Material for}\\
\textbf{"Collective Excitations and Nonequilibrium Phase Transition in Dissipative Fermionic Superfluids"}
\end{center}

\section{Detailed calculations of the time-dependent dissipative BCS theory}
We explain the details of the Hubbard-Stratonovich transformation that is used for the derivation of the time-dependent dissipative BCS theory in the path-integral formalism. The action \eqref{eq_action} in the main text is rewritten as
\begin{align}
S=\int_{-\infty}^\infty dt\Big[&\sum_{\bm k \sigma} \Big( \bar c _{\bm k \sigma +}(i\partial_t-\epsilon_{\bm k}) c_{\bm k \sigma +}-\bar c _{\bm k \sigma -}(i\partial_t-\epsilon_{\bm k}) c_{\bm k \sigma -}\Big)\nonumber\\
&+\sum_{\bm k \bm k'}\left(U\bar c_{\bm{k}\uparrow+}\bar c_{-\bm{k}\downarrow+}c_{-\bm{k}'\downarrow+}c_{\bm{k}'\uparrow+}-U^*\bar c_{\bm{k}\uparrow-}\bar c_{-\bm{k}\downarrow-}c_{-\bm{k}'\downarrow-}c_{\bm{k}'\uparrow-}-i\gamma c_{-\bm{k}\downarrow+}c_{\bm{k}\uparrow+}\bar c_{\bm{k}'\uparrow-}\bar c_{-\bm k'\downarrow-}\right)\Big].
\label{eq_actionS1}
\end{align}
We perform the Hubbard-Stratonovich transformation for each term in the second line of Eq.~\eqref{eq_actionS1} with auxiliary fields $\Delta_\alpha$ ($\alpha=+$, $-$, $\pm$) as
\begin{gather}
iU\bar c_{\bm{k}\uparrow+}\bar c_{-\bm{k}\downarrow+}c_{-\bm{k}'\downarrow+}c_{\bm{k}'\uparrow+}\to-i\Delta_+\bar c_{\bm{k}\uparrow+}\bar c_{-\bm{k}\downarrow+}-i\bar \Delta_+ c_{-\bm{k}\downarrow+}c_{\bm{k}\uparrow+}+\frac{\bar \Delta_+\Delta_+}{iU}\label{eq_U},\\
-iU^*\bar c_{\bm{k}\uparrow-}\bar c_{-\bm{k}\downarrow-}c_{-\bm{k}'\downarrow-}c_{\bm{k}'\uparrow-}\to i\Delta_-\bar c_{\bm{k}\uparrow-}\bar c_{-\bm{k}\downarrow-}+i\bar \Delta_-c_{-\bm{k}\downarrow-}c_{\bm{k}\uparrow-}-\frac{\bar \Delta_-\Delta_-}{iU^*},\\
\gamma c_{-\bm{k}\downarrow+}c_{\bm{k}\uparrow+}\bar c_{\bm{k}'\uparrow-}\bar c_{-\bm k'\downarrow-}\to -\Delta_\pm\bar c_{\bm k\uparrow-}\bar c_{-\bm k \downarrow-}-\bar \Delta_\pm c_{-\bm{k}\downarrow+}c_{\bm{k}\uparrow+}-\frac{\bar \Delta_\pm\Delta_\pm}{\gamma},
\label{eq_Ustar}
\end{gather}
which yield
\begin{align}
S=\int dt\Bigg\{\sum_{\bm k}\Bigg[
\bar \psi_{\bm k +}^t
\left(
\begin{matrix}
i\partial_t-\epsilon_{\bm{k}}&-\Delta_+\\
-\bar \Delta_+ + i\bar \Delta_\pm&-i\partial_t+\epsilon_{\bm k}
\end{matrix}
\right)
\psi_{\bm k +}
-
\bar \psi_{\bm k -}^t
&\left(
\begin{matrix}
i\partial_t-\epsilon_{\bm k}&-\Delta_- -i\Delta_\pm\\
-\bar \Delta_-&-i\partial_t+\epsilon_{\bm k}
\end{matrix}
\right)
\psi_{\bm k -}\Bigg]\notag\\
&+
\frac{\bar \Delta_+\Delta_+}{iU}-\frac{\bar \Delta_-\Delta_-}{iU^*}-\frac{\bar \Delta_\pm\Delta_\pm}{\gamma}\Bigg\},
\label{eq_actionS2}
\end{align}
where $\bar \psi_{\bm k \alpha}=\left(
\begin{matrix}
\bar c_{\bm{k}\uparrow\alpha},&c_{-\bm{k}\downarrow\alpha}
\end{matrix}
\right)^t$ and $\psi_{\bm k \alpha}=\left(
\begin{matrix}
c_{\bm{k}\uparrow\alpha},&\bar c_{-\bm{k}\downarrow\alpha}
\end{matrix}
\right)^t$ ($\alpha=+, -$). Then, from the saddle-point condition $\left\langle \partial S/\partial\Delta_\alpha \right\rangle=\left\langle \partial S/\partial\bar\Delta_\alpha \right\rangle=0$ ($\alpha=+$, $-$, $\pm$), we obtain
\begin{gather}
\Delta_+=-\frac{U}{N_0}\sum_{\bm k}\langle c_{-\bm k\downarrow+}c_{\bm k\uparrow+}\rangle,
\quad\bar \Delta_+=-\frac{U}{N_0}\sum_{\bm k}\langle c_{\bm k\uparrow+}^\dag c_{-\bm k\downarrow+}^\dag\rangle,\label{eq_deltap}\\
\Delta_-=-\frac{U^*}{N_0}\sum_{\bm k}\langle c_{-\bm k\downarrow-}c_{\bm k\uparrow-}\rangle,
\quad\bar \Delta_-=-\frac{U^*}{N_0}\sum_{\bm k}\langle c_{\bm k\uparrow-}^\dag c_{-\bm k\downarrow-}^\dag\rangle,\label{eq_deltam}\\
\Delta_\pm=\frac{\gamma\Delta_+}{U},
\quad\bar \Delta_\pm=\frac{\gamma\bar \Delta_-}{U^*}\label{eq_deltapm},
\end{gather}
where $N_0$ is the number of lattice sites and $\langle\cdots\rangle$ is the expectation value for fixed $\Delta_\alpha$ and $\bar \Delta_\alpha$. Then, we can reduce the number of the auxiliary fields by using $\langle c_{-\bm k\downarrow\alpha}c_{\bm k \uparrow\alpha}\rangle=\mathrm{tr}(c_{-\bm k\downarrow}c_{\bm k \uparrow}\rho)$ ($\alpha=+,-$) and $\mathrm{tr}(A^\dag\rho)=[\mathrm{tr}(A\rho)]^*$ \cite{Kamenev11}, giving
\begin{align}
\Delta_+&=\bar \Delta_-^*,\label{eq_Delta1}\\
\Delta_-&=\bar\Delta_+^*.\label{eq_Delta2}
\end{align}
Finally, the action \eqref{eq_actionS2} is rewritten as
\begin{align}
S=\int dt\sum_{\bm k}\Bigg\{
\bar \psi_{\bm k +}^t
\left(
\begin{matrix}
i\partial_t-\epsilon_{\bm{k}}&-\Delta\\
-\Delta^*&-i\partial_t+\epsilon_{\bm k}
\end{matrix}
\right)
\psi_{\bm k +}
-
\bar \psi_{\bm k -}^t
\left(
\begin{matrix}
i\partial_t-\epsilon_{\bm k}&-\Delta\\
-\Delta^*&-i\partial_t+\epsilon_{\bm k}
\end{matrix}
\right)
\psi_{\bm k -}\Bigg\},
\end{align}
where the superfluid order parameter of the system is given by
\begin{align}
\Delta=-\frac{U}{N_0}\sum_{\bm k}\mathrm{tr}(c_{-\bm k\downarrow}c_{\bm k \uparrow}\rho)\equiv-\frac{U}{N_0}\sum_{\bm k}\langle c_{-\bm k\downarrow}c_{\bm k \uparrow}\rangle.
\end{align}

\section{Operator formalism of the BCS theory for a dissipative superfluid}
Here, we explain the operator formalism of the BCS theory for a dissipative superfluid that leads to Eq.~\eqref{eq_unitary} in the main text. First, we note that an operator $\ket{\psi_+}\bra{\psi_-}$ acting on the Hilbert space of the system can be mapped to a tensor-product state $\ket{\psi_+}\otimes\ket{\psi_-}$ in the doubled Hilbert space $\mathbb H_+\otimes\mathbb H_-$ \cite{Shibata19a, Shibata19b}. Using this mapping, we can rewrite the Liouvillian [see Eq.~\eqref{eq_Lindblad} in the main text for definition] as
\begin{align}
i\mathcal{L}&= H_+ - H_- +i\sum_i \gamma_i(L_{i+}L_{i-}^\dag -\frac{1}{2}L_{i+}^\dag L_{i+}-\frac{1}{2} L_{i-}^\dag L_{i-})\notag\\
&=\mathcal H_+ - \mathcal H_-^* + i\gamma\sum_{\bm k \bm{k}'}c_{-\bm k\downarrow+}c_{\bm k \uparrow+}c_{\bm{k}' \uparrow-}^\dag c_{-\bm{k}' \downarrow-}^\dag,
\label{eq_LiouvilS}
\end{align}
where $L_{i\alpha} = c_{i\downarrow\alpha} c_{i \uparrow\alpha}$, and $c_{i\sigma\alpha}$ ($c_{\bm k\sigma\alpha}$) with $\alpha=+, -$ is the fermion annihilation operator in the real (momentum) space that acts on the Hilbert space $\mathbb{H}_\alpha$. The fermion operator with subscript $+$ ($-$) corresponds to the fermion field in the forward (backward) path in the path-integral formalism. The BCS Hamiltonian equivalent to Eq.~\eqref{eq_Hubbard} is given by
\begin{align}
H_\alpha=\sum_{\bm{k}\sigma}\epsilon_{\bm{k}}c_{\bm{k}\sigma\alpha}^\dagger{c}_{\bm{k}\sigma\alpha}-U_{\mathrm{R}}\sum_{\bm{k}\bm{k}'}c_{\bm{k}\uparrow\alpha}^{\dagger}c_{-\bm{k}\downarrow\alpha}^\dagger{c}_{-\bm{k}'\downarrow\alpha}c_{\bm{k}'\uparrow\alpha},
\end{align}
and $\mathcal H_\alpha$ is defined as
\begin{align}
\mathcal H_\alpha&=H_\alpha-\frac{1}{2}i\gamma \sum_i L_{i\alpha}^\dag L_{i\alpha}\notag\\
&=\sum_{\bm{k}\sigma}\epsilon_{\bm{k}}c_{\bm{k}\sigma\alpha}^\dagger{c}_{\bm{k}\sigma\alpha}-U\sum_{\bm{k}\bm{k}'}c_{\bm{k}\uparrow\alpha}^{\dagger}c_{-\bm{k}\downarrow\alpha}^\dagger{c}_{-\bm{k}'\downarrow\alpha}c_{\bm{k}'\uparrow\alpha}.
\label{eq_HubbardS}
\end{align}
By applying the mean-field approximation to $\mathcal L$, we obtain the mean-field Liouvillian as
\begin{align}
i\mathcal{L}_{\mathrm{MF}}
=&\sum_{\bm k\sigma}\epsilon_{\bm k}c_{\bm k\sigma+}^\dag c_{\bm k\sigma+}+\Delta_+\sum_{\bm k}c_{\bm k\uparrow+}^\dag c_{-\bm k\downarrow+}^\dag+(\bar \Delta_+ -\frac{i\gamma}{U^*}\bar\Delta_-)\sum_{\bm k}c_{-\bm k\downarrow+}c_{\bm k\uparrow+}\nonumber\\
&-\sum_{\bm k\sigma}\epsilon_{\bm k}c_{\bm k\sigma-}^\dag c_{\bm k\sigma-} -(\Delta_- +\frac{i\gamma}{U}\Delta_+)\sum_{\bm k}c_{\bm k\uparrow-}^\dag c_{-\bm k\downarrow-}^\dag-\bar\Delta_-\sum_{\bm k}c_{-\bm k\downarrow-}c_{\bm k\uparrow-}\notag\\
=&\sum_{\bm{k}}
\Psi_{\bm k+}^\dag
\left(
\begin{matrix}
\epsilon_{\bm{k}}&\Delta_+\\
\bar{\Delta}_+-\frac{i\gamma}{U^*}\bar\Delta_-&-\epsilon_{\bm{k}}
\end{matrix}
\right)
\Psi_{\bm k+}
-\sum_{\bm{k}}
\Psi_{\bm k -}^\dag
\left(
\begin{matrix}
\epsilon_{\bm{k}}&\Delta_- +\frac{i\gamma}{U}\Delta_+\\
\bar\Delta_-&-\epsilon_{\bm{k}}
\end{matrix}
\right)
\Psi_{\bm k -},
\label{eq_meanfieldL}
\end{align}
where $\Psi_{\bm k}=\left(\begin{matrix}c_{\bm k \uparrow},&c_{-\bm k \downarrow}^\dag \end{matrix}\right)^t$ is the Nambu spinor. As we can see from Eq.~\eqref{eq_LiouvilS} and Eq.~\eqref{eq_HubbardS}, the Liouvillian is invariant under the U(1) gauge transformations $c_{\bm k\sigma+}\to e^{i\theta}c_{\bm k\sigma+}$ and $c_{\bm k\sigma-}\to e^{i\theta}c_{\bm k\sigma-}$. Moreover, under the exchange of forward and backward operators, $Pc_{\bm k \sigma+}P^{-1}=c_{\bm k \sigma-}$ and $Pc_{\bm k \sigma+}^\dag P^{-1}=c_{\bm k \sigma-}^\dag$ with $P^2=1$, the Liouvillian has the following symmetry 
\begin{align}
P(i\mathcal L)^\dag P^{-1}=-i\mathcal{L}.
\end{align}
By imposing the same symmetry on the mean-field Liouvillian as $P(i\mathcal{L}_{\mathrm{MF}})^\dag P^{-1}=-i\mathcal{L}_{\mathrm{MF}}$, we obtain the relations for the order parameters as
\begin{align}
\Delta_+^*&=\bar \Delta_-,\\
\Delta_-^*&=\bar\Delta_+,
\end{align}
which coincide with those obtained in the path-integral formalism [see Eqs. \eqref{eq_Delta1} and \eqref{eq_Delta2}]. Finally, by rewriting the superfluid order parameter $\Delta_+$ as
\begin{align}
\Delta=-\frac{U}{N_0}\sum_{\bm k}\mathrm{tr}(c_{-\bm k\downarrow}c_{\bm k \uparrow}\rho)\equiv-\frac{U}{N_0}\sum_{\bm k}\langle c_{-\bm k\downarrow}c_{\bm k \uparrow}\rangle,
\end{align}
we obtain the equations for the density matrix as
\begin{gather}
\dot{\rho}=-i[H_{\mathrm{eff}},\rho],
\label{eq_unitaryS}\\
H_{\mathrm{eff}}=\sum_{\bm{k}}
\Psi_{\bm k}^\dag
\left(
\begin{matrix}
\epsilon_{\bm{k}}&\Delta\\
\Delta^*&-\epsilon_{\bm{k}}
\end{matrix}
\right)
\Psi_{\bm k},
\label{eq_HamiltonianEffS}
\end{gather}
which are the same as Eqs.~\eqref{eq_unitary} and \eqref{eq_HamiltonianEff} in the main text.

\section{Generalization of the Bogoliubov-de Gennes analysis with a time-dependent BCS state}
We explain that Eq.~\eqref{eq_unitary} (Eq.~\eqref{eq_bloch} in the psedouspin representation) in the main text is equivalent to the Bogoliubov-de Gennes equation with a time-dependent BCS state \cite{Barankov06A, Volkov74}, which describes the unitary evolution of the density matrix.

We introduce the time-dependent BCS state of the effective Hamiltonian $H_{\mathrm{eff}}=\sum_{\bm{k}}
\Psi_{\bm k}^\dag
\left(
\begin{matrix}
\epsilon_{\bm{k}}&\Delta\\
\Delta^*&-\epsilon_{\bm{k}}
\end{matrix}
\right)
\Psi_{\bm k}$
as follows:
\begin{gather}
|\Psi_{\mathrm{BCS}}(t)\rangle=\prod_{\bm k} (u_{\bm k}(t) + v_{\bm k}(t) c_{\bm k \uparrow}^\dag c_{-\bm k \downarrow}^\dag)|0\rangle\label{eq_BCS},\\
|u_{\bm k}|^2+|v_{\bm k}|^2=1,
\end{gather}
where $|0\rangle$ is the vacuum of fermions. Here, the superfluid order parameter $\Delta$ is rewritten as
\begin{align}
\Delta=-\frac{U}{N_0}\sum_{\bm k}\langle c_{-\bm k \downarrow} c_{\bm k \uparrow}\rangle=-\frac{U}{N_0}\sum_{\bm k} u_{\bm k}^*(t)v_{\bm k}(t).
\end{align}
Suppose that the density matrix is given by 
\begin{align}
\rho(t)=\ket{\Psi_{\mathrm{BCS}}(t)}\bra{\Psi_{\mathrm{BCS}}(t)}.
\label{eq_rhoBCS}
\end{align}
Then, the time-evolution equation
\begin{align}
\frac{d\rho(t)}{dt}=-i[H_{\mathrm{eff}},\rho(t)]
\label{eq_unitaryS1}
\end{align}
[Eq.~\eqref{eq_unitary} in the main text] is equivalent to the Bogoliubov-de Gennes equation with the time-dependent BCS state
\begin{align}
i\partial_t
\left(
\begin{matrix}
u_{\bm k}\\
v_{\bm k}
\end{matrix}
\right)
=
\left(
\begin{matrix}
-\epsilon_{\bm k}&\Delta^*\\
\Delta&\epsilon_{\bm k}
\end{matrix}
\right)
\left(
\begin{matrix}
u_{\bm k}\\
v_{\bm k}
\end{matrix}
\right).
\label{eq_BogoliubovS}
\end{align}
By defining $f_{\bm k}(t)$ and $g_{\bm k}(t)$ as
\begin{gather}
f_{\bm k}=u_{\bm k}^* v_{\bm k},\\
g_{\bm k}=\frac{1}{2}(|u_{\bm k}|^2 - |v_{\bm k}|^2),
\end{gather}
Eq.~\eqref{eq_BogoliubovS} is rewritten as
\begin{gather}
\frac{d f_{\bm k}}{dt}=-2i\epsilon_{\bm k}f_{\bm k}-2i\Delta g_{\bm k},\\
\frac{d g_{\bm k}}{dt}=i\Delta f_{\bm k}^*-i\Delta^*f_{\bm k}.
\end{gather}
These equations take the same forms as those for closed systems \cite{Barankov06A, Volkov74}. Finally, by defining the psedouspins as
\begin{gather}
f_{\bm k}=\sigma_{\bm k}^x -i \sigma_{\bm k}^y,\\
g_{\bm k}=-\sigma_{\bm k}^z,
\end{gather}
we obtain the same Bloch equation as discussed in the main text:
\begin{gather}
\frac{d\bm \sigma_{\bm k}}{dt}
=2\bm b_{\bm k}\times\bm \sigma_{\bm k},\label{eq_blochS}\\
\bm b_{\bm k}=\left(\begin{matrix}\mathrm{Re}\Delta,&-\mathrm{Im}\Delta,&\epsilon_{\bm k}\end{matrix}\right).
\end{gather}
We note that the dynamics described by Eq.~\eqref{eq_unitaryS1} conserves the purity $\mathrm{tr}[\rho^2]$ as
\begin{align}
\frac{d \mathrm{tr}[\rho^2]}{d t}=2\mathrm{tr}\left[\rho \frac{d\rho}{dt}\right]=-2i\mathrm{tr}(\rho[H_{\mathrm{eff}}, \rho])=0.
\end{align}

In general, the purity should decrease during the time evolution described by the quantum master equation [Eq.~\eqref{eq_Lindblad} in the main text]. This fact is consistent with the time-dependent BCS ansatz \eqref{eq_BCS} as follows. Since $|\Psi_{\mathrm{BCS}}(t)\rangle$ can be expanded in terms of $N$-particle states
\begin{align}
|\Psi_N\rangle=\sum_{\bm k_1\cdots\bm k_{N/2}}a_{\bm k_1}\cdots a_{\bm k_{N/2}}c_{\bm k_1 \uparrow}^\dag c_{-\bm k_1 \downarrow}^\dag \cdots c_{\bm k_{N/2} \uparrow}^\dag c_{-\bm k_{N/2} \downarrow}^\dag|0\rangle
\end{align}
as
\begin{align}
|\Psi_{\mathrm{BCS}}(t)\rangle=\sum_N c_N|\Psi_N\rangle,
\label{eq_pureS}
\end{align}
the density matrix is written as
\begin{align}
\rho&=|\Psi_{\mathrm{BCS}}(t)\rangle\langle\Psi_{\mathrm{BCS}}(t)|\notag\\
&=\sum_N |c_N|^2|\Psi_N\rangle\langle\Psi_N| + \sum_{N\neq N^\prime} c_{N^\prime}^* c_N|\Psi_N\rangle\langle\Psi_{N^\prime}|.
\label{eq_rhoS}
\end{align}
Then, for a gauge-invariant observable $\mathcal O$, its expectation value is given by
\begin{align}
\langle\mathcal O\rangle\equiv\mathrm{tr}[\mathcal O \rho]=\sum_N|c_N|^2\langle\Psi_N|\mathcal O|\Psi_N\rangle=\mathrm{tr}[\mathcal O \rho^\prime],
\end{align}
where
\begin{align}
\rho^\prime=\sum_N |c_N|^2|\Psi_N\rangle\langle\Psi_N|
\label{eq_rhomixed}
\end{align}
is a mixed state of different particle numbers. Therefore, concerning gauge-invariant observables, the time-dependent BCS state \eqref{eq_rhoBCS} is indistinguishable from the mixed state \eqref{eq_rhomixed} with $\mathrm{tr}[\rho^{\prime 2}] < 1$. Since any physically observable quantity should be gauge invariant, the time-dependent BCS ansatz \eqref{eq_BCS} can describe the time evolution of the density matrix consistently with the quantum master equation \eqref{eq_Lindblad}.

\section{Dynamics of pseudospins after switch-on of dissipation}
\begin{figure}[h]
\includegraphics[width=15cm]{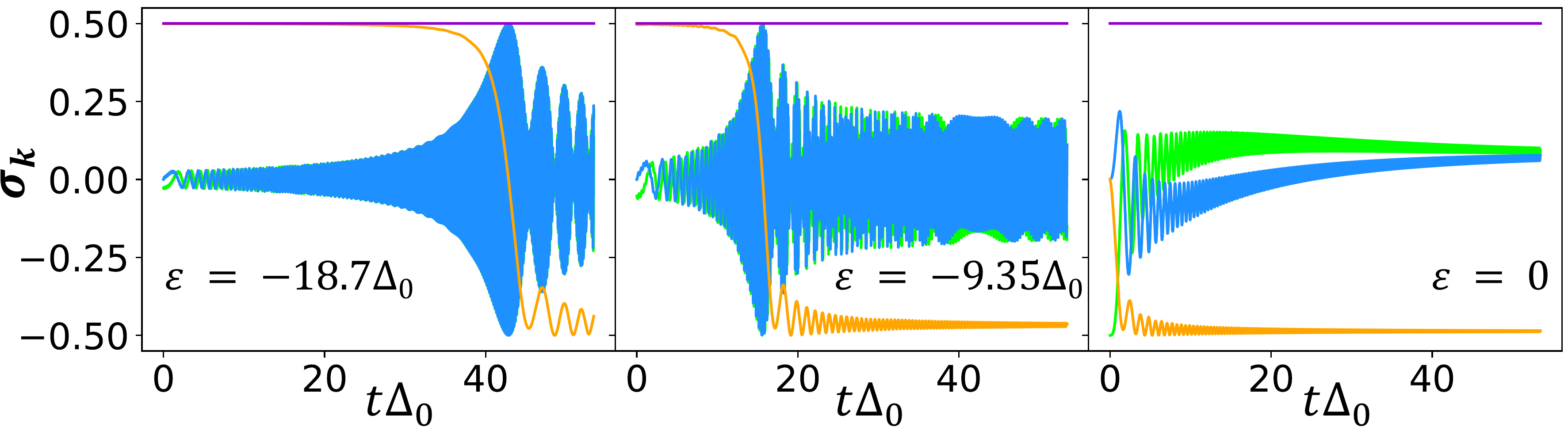}
\caption{Dynamics of pseudospins [$\sigma_{\bm k}^x$ (light green), $\sigma_{\bm k}^y$ (blue), $\sigma_{\bm k}^z$ (orange), $|\bm \sigma_{\bm k}|$ (violet)] after the atom loss with $\gamma=2.81\Delta_0$ is switched on for the initial state with $U_{\mathrm R}=12.2\Delta_0$ and the Fermi energy $\epsilon_{\mathrm F}=0$ for $\epsilon=-18.7\Delta_0$ (left), $-9.35\Delta_0$ (center), and $0$ (right), where $\epsilon$ is the single-particle energy in the band $-23.4\Delta_0\le\epsilon\le23.4\Delta_0$.}
\label{fig_psedou}
\end{figure}

The physical origin of the chirping of the U(1) phase can be understood from the pseudospin picture. As shown in Fig.~\ref{fig_psedou}, when the sign of $\sigma_{\bm k}^z$ changes from positive to negative, the magnitudes of $\sigma_{\bm k}^x$ and $\sigma_{\bm k}^y$ increase due to the norm conservation of pseudospins. This indicates that the Cooper-pair amplitude at specific momenta rapidly changes when atoms at those momenta are lost from the system. Since Cooper pairs are formed near the Fermi surface, a loss of Cooper pairs leads to a downward shift of the Fermi level.

To see the effect of the dynamics of pseudospins on the collective phase mode, we calculate the angular velocity of the order parameter. From Eq.~(11) in the main text, the real and imaginary parts of the order parameter are written as
\begin{align}
|\Delta|\cos\theta =\mathrm{Re}\Delta = -U_1\sum_{\bm k} \sigma_{\bm k}^x - \frac{\gamma}{2}\sum_{\bm k} \sigma_{\bm k}^y,\label{eq_reOPS}\\
|\Delta|\sin\theta =\mathrm{Im}\Delta = - \frac{\gamma}{2}\sum_{\bm k} \sigma_{\bm k}^x + U_1\sum_{\bm k} \sigma_{\bm k}^y. \label{eq_imOPS}
\end{align}
By differentiating Eqs.~\eqref{eq_reOPS} and \eqref{eq_imOPS} with respect to time, we obtain
\begin{align}
|\Delta|^2\frac{d\theta}{dt} = \left(U_1 \mathrm{Im}\Delta - \frac{\gamma}{2}\mathrm{Re}\Delta\right)\sum_{\bm k}\frac{d\sigma_{\bm k}^x}{dt} + \left(\frac{\gamma}{2}\mathrm{Im}\Delta + U_1\mathrm{Re}\Delta\right)\sum_{\bm k}\frac{d\sigma_{\bm k}^y}{dt}.
\label{eq_pseudoderivative}
\end{align}
Then, by substituting the Bloch equation \eqref{eq_blochS} into Eq.~\eqref{eq_pseudoderivative}, we arrive at
\begin{align}
\frac{d\theta}{dt} = U_{\mathrm{R}}(1-\frac{N}{N_0})-\frac{2|U|^2}{|\Delta|^2N_0^2}\sum_{\substack {\bm k \bm k' \\ \alpha=x,y}}\sigma_{\bm k}^\alpha\cdot\epsilon_{\bm k'}\sigma_{\bm k'}^\alpha,
\label{eq_thetadot}
\end{align}
in which the last term increases due to the shift of the Fermi level, leading to chirping of the U(1) phase. Here, we note that the first term on the right-hand side of Eq.~\eqref{eq_thetadot} also increases as the particle number decreases; however, it is much smaller than the second term.

\section{Dynamics after sudden change of both the interaction and the dissipation}
\begin{figure}[h]
\includegraphics[width=15cm]{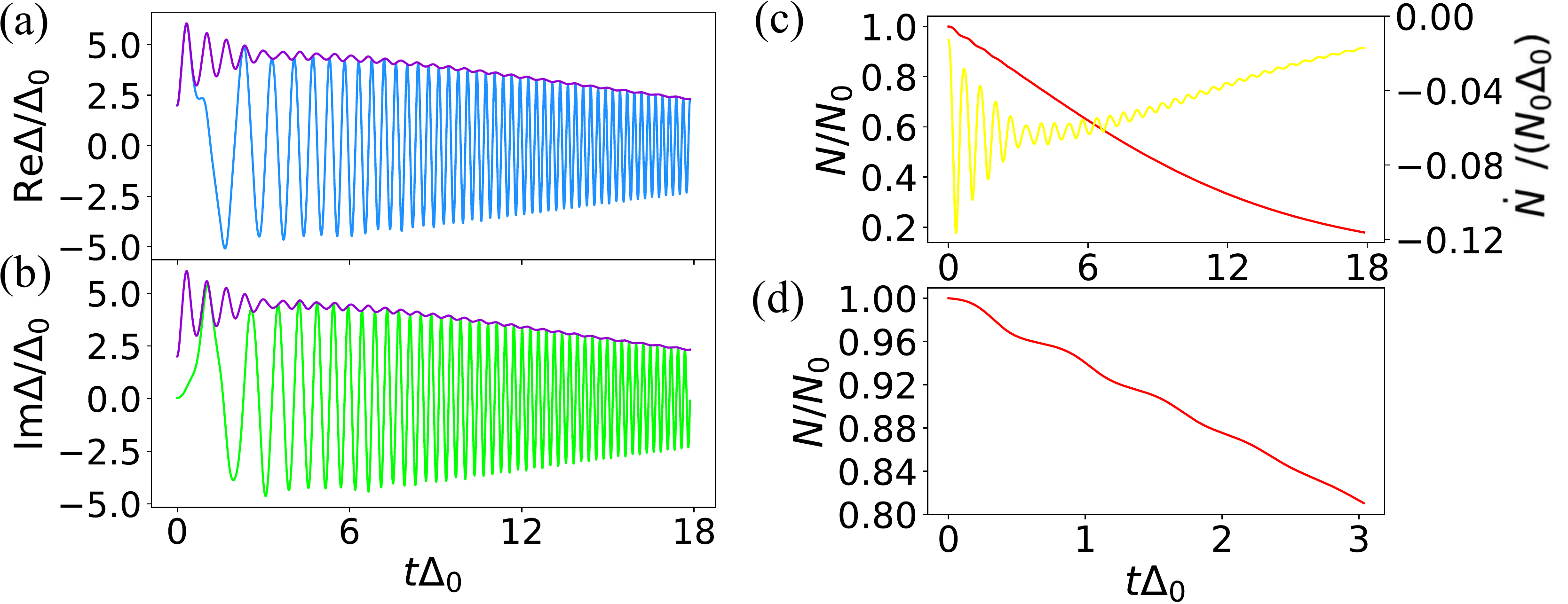}
\caption{Dynamics of a superfluid after the interaction and the atom loss are suddenly changed. (a) Real parts (light green) and imaginary parts (blue) of the order parameter. The amplitudes of the order parameter are shown as violet curves in both figures. (c) The particle number of the system normalized by the initial particle number $N_0$ (red) and the rate of change in particle number (yellow). (d) An enlarged view of the particle number near $t=0$. The parameters are suddenly changed from $U_{\mathrm R}=8.4\Delta_0$ to $U=(16.8+0.45i)\Delta_0$ at $t=0$ and the bandwidth is set to $W=28\Delta_0$.}
\label{fig_intloss}
\end{figure}

Figure \ref{fig_intloss} shows the dynamics after the dissipation $\gamma$ is introduced at $t=0$ and the interaction strength is simultaneously changed from $U_{\mathrm R}=8.4\Delta_0$ to $U_{\mathrm R}=16.8\Delta_0$. In Figs.~\ref{fig_intloss}(a) and (b), we see an amplitude oscillation larger than that in the loss quench dynamics shown in Figs.~\ref{fig_lossquench}(a) and (b) in the main text. This behavior is due to the fact that a change in the real part of $U$ causes a large initial shift of the amplitude of the order parameter [see Fig.~\ref{fig_Schematic}(a) in the main text]. The U(1) phase rotates with an increasing angular velocity due to dissipation as in Figs.~\ref{fig_intloss}(a) and (b). The amplitude oscillation of the order parameter can be detected through monitoring of the particle number and from Eq.~\eqref{eq_Ndt} in the main text. As shown in Figs.~\ref{fig_intloss}(c) and (d), the amplitude of the particle-number oscillation is a few percent of the initial particle number, which can be detected with current experimental techniques \cite{Roati15}. Since the oscillation in the particle number cannot appear in isolated systems, the dissipation-induced dynamics can be used as a unique signature for the amplitude mode of the superfluid order parameters.

\section{A simplified model for understanding the nonequilibrium phase transition}
Here, we explain how the nonequilibrium phase transition associated with the vanishing dc Josephson current can be understood by considering a simplified model of the Josephson junction with particle loss. We consider two fermionic superfuids coupled via a Josephson junction, where two-body loss is introduced to one of them, as shown in Fig.~\ref{fig_Schematic}(b) in the main text. The Josephson current flowing between the two systems is given by
\begin{align}
I=\frac{1}{N_0}\frac{dN_1}{dt}=-I_0\sin(\Delta\theta),
\end{align}
where, from Eq.~\eqref{eq_JosephsonCurrent} in the main text, $\Delta\theta=\theta_2-\theta_1$, $I_0=4V|\Delta_1||\Delta_2|/U_{\mathrm R}|U|$, and we have neglected the phase shift $\delta$ because $\delta\ll\Delta\theta$. Taking into account the particle loss [see Eq.~\eqref{eq_Ndt} in the main text], we obtain the rate of change in the particle number of system 1 and that of system 2 as
\begin{gather}
\frac{1}{N_0}\frac{dN_1}{dt}=-I_0\sin(\Delta\theta),\label{eq_N1S}\\
\frac{1}{N_0}\frac{dN_2}{dt}=I_0\sin(\Delta\theta)-\frac{2\gamma|\Delta_2|^2}{|U|^2}.\label{eq_N2S}
\end{gather}
We assume that the time evolution of the phase difference $\Delta\theta$ is given by the effective chemical-potential difference $\Delta\mu_{\mathrm {eff}}$ between the two systems as
\begin{align}
\frac{d\Delta\theta}{dt}=-2\Delta\mu_{\mathrm{eff}}=-\frac{W}{N_0}(N_2-N_1),\label{eq_thetaS}
\end{align}
where $W$ is a bandwidth and we assume a constant density of states for simplicity (we can also understand this equation from the phenomenological time-dependent Ginzburg-Landau theory, which is explained in the last part of this section). We obtain the equation of motion for $\Delta\theta$ by using Eqs.~\eqref{eq_N1S}, \eqref{eq_N2S}, and \eqref{eq_thetaS} as
\begin{align}
\frac{d^2\Delta\theta}{dt^2}=-2WI_0\sin(\Delta\theta)+\frac{2\gamma W|\Delta_2|^2}{|U|^2}.
\label{eq_Deltatheta}
\end{align}
This system is regarded as a Josephson junction with shunt resistance $R=+\infty$, capacitance $C=1/2W$ and an external force $F=\gamma|\Delta_2|^2/|U|^2$, which is described as
\begin{align}
C\frac{d^2\Delta\theta}{dt^2}+\frac{1}{R}\frac{d\Delta\theta}{dt}+I_0\sin(\Delta\theta)=F.
\label{eq_washboard}
\end{align}
That is, the time evolution of $\Delta\theta$ is equivalent to that of a particle moving in a washboard potential
\begin{align}
V_{\mathrm{wash}}=-2WI_0\cos(\Delta\theta)-\frac{2\gamma W|\Delta_2|^2\Delta\theta}{|U|^2}.
\end{align}
The condition for the extremum of $V_{\mathrm {wash}}$ is given by $dV_{\mathrm {wash}}/d\Delta\theta=0$, giving
\begin{align}
\sin(\Delta\theta)=\frac{\gamma|\Delta_2|^2}{I_0|U|^2}.
\end{align}
The solution to this equation does not exist for $\gamma|\Delta_2|^2/I_0|U|^2> 1$ and the time evolution of $\Delta\theta$ becomes unstable. If we assume $|\Delta_1|\simeq|\Delta_2|$ when the time evolution is sufficiently slow, we obtain the critical strength of the atom loss as
\begin{align}
\gamma_c\simeq4V,
\end{align}
which is of the same order of magnitude as that in the main text ($\gamma_c\simeq3V$). Thus, the system exhibits a dynamical phase transition from the state in which $\Delta\theta$ oscillates around an extremum of $V_{\mathrm{wash}}$ for $\gamma<\gamma_c$ to the state in which $\Delta\theta$ slips down the washboard potential for $\gamma>\gamma_c$. Thus, the dynamical phase transition caused by the particle loss is the one between a trapped state and a running state. We note that the particle loss $\gamma$ acts as an external force $F$ rather than friction $R$ in Eq.~\eqref{eq_washboard}. The loss-induced dynamical phase transition occurs spontaneously without any external fields, and has an essentially different origin from the localization-delocalization transition induced by friction $R$ \cite{Caldeira81, Schdmit83, Guinea85}.

\begin{figure}[h]
\includegraphics[width=15cm]{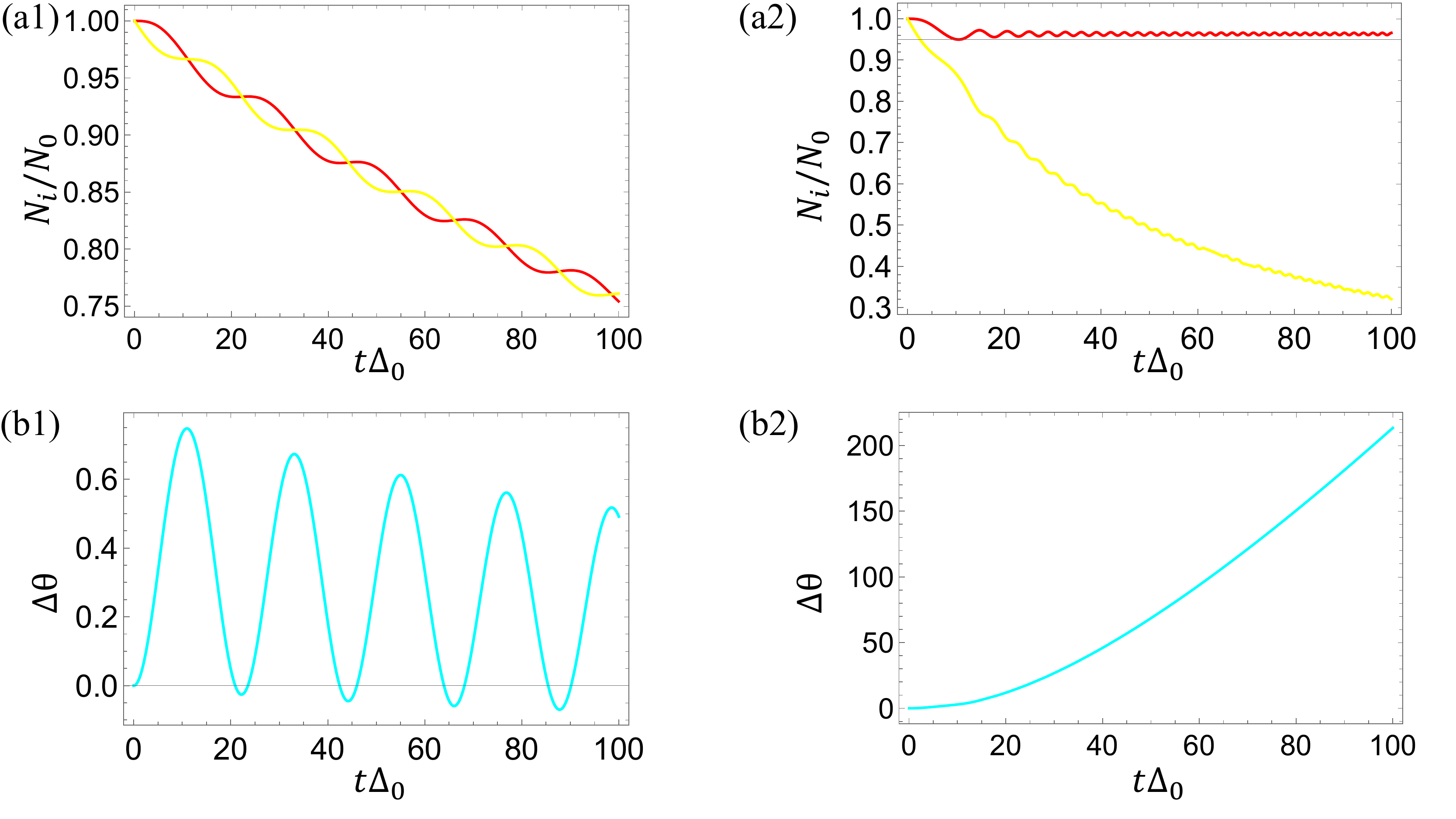}
\caption{Numerical solution of $N_1/N_0$ (red), $N_2/N_0$ (yellow), and $\Delta\theta$ (light blue) in Eqs.~\eqref{eq_N1Sa}, \eqref{eq_N2Sa}, and \eqref{eq_Deltatheta1} with $W=5.11\Delta_0$ and $I_0=0.009\Delta_0$. The loss rate $\kappa_2$ is set to $\kappa_2=0.006\Delta_0$ in (a1) and (b1), and $\kappa_2=0.02\Delta_0$ in (a2) and (b2).}
\label{fig_phenomenology}
\end{figure}

These features are also obtained from a phenomenological introduction of two-body loss \cite{Syassen08, Garcia09}, under which the rate of change in the particle number of system 1 and that of system 2 are given by 
\begin{gather}
\frac{1}{N_0}\frac{dN_1}{dt}=-I_0\sin(\Delta\theta),\label{eq_N1Sa}\\
\frac{1}{N_0}\frac{dN_2}{dt}=I_0\sin(\Delta\theta)-\kappa_2\left(\frac{N_2}{N_0}\right)^2,\label{eq_N2Sa}
\end{gather}
where $\kappa_2$ is the two-body loss rate. By using Eqs.~\eqref{eq_thetaS}, \eqref{eq_N1Sa}, and \eqref{eq_N2Sa}, we obtain the equation of motion for $\Delta\theta$ as
\begin{align}
\frac{d^2\Delta\theta}{dt^2}=-2WI_0\sin(\Delta\theta)+W\kappa_2\left(\frac{N_2}{N_0}\right)^2.
\label{eq_Deltatheta1}
\end{align}
This system is regarded as a Josephson junction with resistance $R=+\infty$, capacitance $C=1/2W$ and an external force $F=\frac{\kappa_2}{2}\left(\frac{N_2}{N_0}\right)^2$. As the system has the washboard potential
\begin{align}
V_{\mathrm{wash}}=-2WI_0\cos(\Delta\theta)-W\kappa_2\left(\frac{N_2}{N_0}\right)^2\Delta\theta,
\end{align}
the condition for the extremal of $V_{\mathrm{wash}}$ is given by
\begin{align}
\sin(\Delta\theta)=\frac{\kappa_2}{2I_0}\left(\frac{N_2}{N_0}\right)^2.
\end{align}
If we assume that the variation of the parameters is sufficiently slow and approximate them as constant, the critical strength of the loss rate where the solution of $\Delta\theta$ becomes unstable is given by
\begin{align}
\kappa_{2c}\simeq2I_0.
\end{align}
We can numerically solve Eq.~\eqref{eq_N1Sa}, \eqref{eq_N2Sa}, and \eqref{eq_Deltatheta1}, and the results are shown in Fig.~\ref{fig_phenomenology}. We see that the results shown in Fig.~\ref{fig_phenomenology}(a1) and (b1) [Fig.~\ref{fig_phenomenology}(a2) and (b2)] are qualitatively the same as those in Fig.~\ref{fig_multiquench}(c1) and (d1) [Fig.~\ref{fig_multiquench}(c2) and (d2)] in the main text, respectively.

The time-evolution equation of the phase [Eq.~\eqref{eq_thetaS}] can be understood as a consequence of the gauge transformation ($\Delta_i(t)=\exp(-2i\int_0^tdt\mu_{i \mathrm{eff}}(t))|\Delta_i(t)|$), which reflects the conjugate nature of the particle number and the phase. From a more phenomenological point of view, Eq. (S47) can be understood using the time-dependent Ginzburg-Landau (GL) theory. Here we note that, strictly speaking, the GL theory cannot be applied to zero-temperature superfluids considered in our study, since the Taylor expansion of the free energy requires that the system should be close to the transition temperature. In nonequilibrium situations, the time-dependent GL theory cannot be applied to gapped superfluids and nonadiabatic regimes, since the time scale of the order-parameter dynamics becomes shorter than the lifetime of quasiparticles and quasiparticle contributions cannot be neglected \cite{Barankov04, Tsuji15}. Moreover, for the application of the time-dependent GL theory, it is required that the deviation from equilibrium is sufficiently small, which cannot be satisfied in our dissipative superfluids which are driven far from equilibrium. However, we assume below that the GL theory could be phenomenologically used to investigate the dynamics of the dissipative superfluids. We start from the phenomenological GL equation of the superfluid condensate, which is expanded in a series of the order parameter
\begin{align}
-\Gamma\left(\frac{\partial \Delta}{\partial t} + \frac{2ie\varphi}{\hbar}\Delta\right)=a\Delta+b|\Delta|^2\Delta,
\end{align}
where $\Gamma$ is a positive constant, $\varphi$ is the scalar potential, $e$ is the charge, and we have assumed that the order parameter is spatially uniform. Here, we have introduced complex-valued coefficients $a$ and $b$, which might describe the effect of the atom loss. However, we remark on the validity of this treatment below. In our study, the scalar potential does not exist. Instead, we introduce an effective chemical potential $\mu_{\mathrm{eff}}$ that is determined from the total particle number of the system as follows:
\begin{align}
-\Gamma\left(\frac{\partial \Delta}{\partial t} + 2i\mu_{\mathrm{eff}}\Delta\right)=a\Delta+b|\Delta|^2\Delta.
\end{align}
By rewriting the equation in terms of the phase $\theta$ and the amplitude $|\Delta|$ of the order parameter (the latter is proportional to the square root of the superfluid density), we obtain the equation of motion for the phase as
\begin{align}
\frac{\partial\theta}{\partial t}=-2\mu_{\mathrm{eff}}-\frac{\mathrm{Im}(a)+\mathrm{Im}(b)|\Delta|^2}{\Gamma}.
\label{eq_ThetaGL}
\end{align}
For two superfluids connected via a Josephson junction, we similarly obtain
\begin{align}
\frac{\partial(\theta_2-\theta_1)}{\partial t}=-2(\mu_{2\mathrm{eff}}-\mu_{1\mathrm{eff}})-\frac{\mathrm{Im}(a_2)+\mathrm{Im}(b_2)|\Delta_2|^2}{\Gamma_2},
\label{eq_DeltaThetaS}
\end{align}
where the subscript 1 and 2 label the two superfluids, and we have introduced the effect of loss to superfluid 2. If the coefficients $a$ and $b$ are real, we arrive at Eq.~\eqref{eq_thetaS}:
\begin{align}
\frac{\partial\Delta\theta}{\partial t}=-2\Delta\mu_{\mathrm{eff}}.
\label{eq_thetaS2}
\end{align}
We note, however, that Eqs.~\eqref{eq_N1S} and \eqref{eq_N2S} cannot be obtained from the GL theory, since $N_1$ and $N_2$ are the total particle numbers of the two superfluids, which are not directly related to the amplitude of the order parameter for dissipative superfluids. In the case of complex coefficients $a$ and $b$, one might at first sight think that the effect of loss can be described by regarding the second term in Eq.~\eqref{eq_DeltaThetaS} as an effective change in the chemical potential, which reads $\Delta\mu\simeq W(N-N_0)/N_0\propto|\Delta|^2-|\Delta_0|^2$. However, this result of the phenomenological treatment of the time-dependent GL theory has a serious problem. To see this, let us differentiate this equation with respect to time as $dN/dt\propto |\Delta|d|\Delta|/dt$. As the amplitude of the order parameter oscillates as shown in Fig.~2(a) in the main text, the equation $dN/dt\propto|\Delta|d|\Delta|/dt$ indicates that the total particle number of the system can increase since the oscillating part $d|\Delta|/dt$ can take a positive value. Such an increase of the particle number is unphysical since our system has no particle gain. As the correct equation obtained from the microscopic theory is $dN/dt\propto-|\Delta|^2$ [Eq.~\eqref{eq_Ndt} in the main text], which indicates that the total particle number of the system monotonically decreases, Eq.~\eqref{eq_DeltaThetaS} does not correctly describe the dynamics associated with the loss in the total particle number. Thus, it is highly nontrivial to consistently describe the dynamic evolution of the order parameter, collective excitations, and the dynamical phase transition caused by the particle loss in dissipative fermionic superfluids on the basis of the phenomenological GL theory.

If we wish to derive the time-dependent GL equation from the formalism based on the closed-time-contour path integral, we should start from a generating functional like Eq. (3) in the main text, and introduce the order parameter by performing the Hubbard-Stratonovich transformation. Then, by integrating out the fermionic degrees of freedom, we arrive at the action with respect to the order parameter. By expanding the exact action $S(\Delta,\Delta^*)$ around its extremal value $S(\Delta_0,\Delta_0^*)$, where $\Delta_0$ is usually the equilibrium value that satisfies the BCS gap equation, the time-dependent GL equation is obtained from a saddle-point equation $\partial S(\Delta,\Delta^* )/\partial\Delta^*(r,t)=0$ \cite{Stoof93}. However, we have to pay careful attention to the condition that is assumed in the standard derivation: the deviation of the order parameter $\Delta$ around its extremal value $\Delta_0$ should be small enough for the Taylor expansion to be justified. We also note that the standard derivation assumes that the total particle number of the system does not change. In contrast, as the particle number significantly changes in our system, the phase of the order parameter rotates and largely deviates from the stationary value as shown in Fig. 2 in the main text. Thus, a time-dependent GL theory that fully incorporates the change in particle number needs highly nontrivial consideration, which deserves further study.

\end{document}